\documentclass{jfm}

\usepackage{graphicx}
\usepackage{natbib}

\ifCUPmtlplainloaded \else
  \checkfont{eurm10}
  \iffontfound
    \IfFileExists{upmath.sty}
      {\typeout{^^JFound AMS Euler Roman fonts on the system,
                   using the 'upmath' package.^^J}%
       \usepackage{upmath}}
      {\typeout{^^JFound AMS Euler Roman fonts on the system, but you
                   dont seem to have the}%
       \typeout{'upmath' package installed. JFM.cls can take advantage
                 of these fonts,^^Jif you use 'upmath' package.^^J}%
      }
  \else
  \fi
\fi

\ifCUPmtlplainloaded \else
  \checkfont{msam10}
  \iffontfound
    \IfFileExists{amssymb.sty}
      {\typeout{^^JFound AMS Symbol fonts on the system, using the
                'amssymb' package.^^J}%
       \usepackage{amssymb}%

      }{}
  \fi
\fi

\ifCUPmtlplainloaded \else
  \IfFileExists{amsbsy.sty}
    {\typeout{^^JFound the 'amsbsy' package on the system, using it.^^J}%
     \usepackage{amsbsy}}
    {}
\fi

\newcommand\Pran{\mbox{\textit{Pr}}} 

\newsavebox{\astrutbox}
\sbox{\astrutbox}{\rule[-5pt]{0pt}{20pt}}

\newcommand\be{\begin{equation}}
\newcommand\ee{\end{equation}}
\newcommand\bes{\begin{equation*}}
\newcommand\ees{\end{equation*}}

\newcommand{\dv}[2]{\frac{d#1}{d#2}}

\title[Closed-form shock solutions]{Closed-form shock solutions}

\author[B. M. Johnson]%
{B. M. Johnson$^1$%
  \thanks{Email address for correspondence: johnson359@llnl.gov}}

\affiliation{$^1$Lawrence Livermore National Laboratory, Livermore,
CA 94550, USA}

\pubyear{2010}
\volume{650}
\pagerange{119--126}
\date{13 December 2013; revised 10 February 2014; accepted 24 February 2014}
\begin{document}

\maketitle

\begin{abstract}
It is shown here that a subset of the implicit analytical shock solutions discovered by Becker and by Johnson can be inverted, yielding several exact closed-form solutions of the one-dimensional compressible Navier-Stokes equations for an ideal gas. For a constant dynamic viscosity and thermal conductivity, and at particular values of the shock Mach number, the velocity can be expressed in terms of a polynomial root. For a constant kinematic viscosity, independent of Mach number, the velocity can be expressed in terms of a hyperbolic tangent function. The remaining fluid variables are related to the velocity through simple algebraic expressions. The solutions derived here make excellent verification tests for numerical algorithms, since no source terms in the evolution equations are approximated, and the closed-form expressions are straightforward to implement. The solutions are also of some academic interest as they may provide insight into the non-linear character of the Navier-Stokes equations and may stimulate further analytical developments.
\end{abstract}


\section{Introduction}

One of the few known non-linear analytical solutions to the equations of fluid dynamics was discovered by \cite{Becker22} and subsequently analyzed by \cite{Thomas44}, \cite{Morduchow49}, \cite{Hayes60} and \cite{Iannelli13}. It captures the physical profile of shock fronts in ideal gases, and although it requires some restrictive assumptions (a steady state, one planar dimension, constant dynamic viscosity, an ideal gas equation of state and a constant Prandtl number $\Pran$ of $3/4$), the solution is exact in the sense that no source terms in the (one-dimensional) evolution equations are neglected or approximated. Analogous solutions were discovered by \cite{Johnson13} in the limit of both large and small $\Pran$. These solutions provide a useful framework for verifying numerical algorithms used to solve the Navier-Stokes equations. A drawback, however, from the perspective of both physical intuition and numerical implementation, is that the solutions are implicit, i.e., they are solutions for $x(v)$ rather than closed-form expressions for $v(x)$ ($x$ here is the spatial dimension in which the shock propagates and $v$ is the velocity magnitude).

It is shown here that some of these implicit solutions can be inverted for particular values of the shock Mach number, yielding closed-form expressions for the fluid velocity as a function of position. In particular, for rational values of the shock compression ratio, Becker's implicit expression is a polynomial in $v(x)$. Expressions for the polynomial root relevant to a shock are provided up to a compression ratio of four. Polynomial solutions also exist in both the large- and small-$\Pran$ limits under the assumption of either a constant dynamic viscosity or constant thermal conductivity, and expressions are provided for these as well. Under the assumption of a constant kinematic (rather than dynamic) viscosity, the solution for $v(x)$ takes the particularly simple form of a hyperbolic tangent function; this solution is valid at any Mach number and for both $\Pran \rightarrow \infty$ and $\Pran \rightarrow 3/4$.

An overview of the equations to be solved is given in \S\ref{sec:equations}, the solutions are given in \S\ref{sec:solutions}, and a summary is given in \S\ref{sec:summary}.

\section{Equations}\label{sec:equations}

In one planar dimension and a steady state, the compressible Navier-Stokes equations reduce to the following ordinary differential equations: 
\be\label{eq:veq}
\frac{4\mu}{3m_0} v\dv{v}{x} = v^2 + \frac{\gamma - 1}{\gamma} h - \frac{\gamma+1}{2 \gamma}\left(v_0 + v_1\right) v,
\ee
\be\label{eq:heq}
\frac{\kappa}{m_0 C_p} \dv{h}{x} = \frac{h}{\gamma} - \frac{v^2}{2} + \frac{\gamma+1}{2 \gamma}\left(v_0 + v_1\right) v - \frac{\gamma+1}{\gamma - 1}\frac{v_0v_1}{2},
\ee
where $\rho$ is the mass density, $h = e + p/\rho$ is the fluid enthalpy, $p$ is the pressure, $e$ is the internal energy, $\mu$ is the dynamic viscosity (in the limit of negligible bulk viscosity; otherwise $\mu$ is the sum of the dynamic viscosity and $3/4$ of the bulk viscosity), $\kappa$ is the thermal conductivity and $m_0 = \rho v = \rho_0 v_0$ is the mass flux \citep{Becker22,Zel'dovich02,Johnson13}. It has been assumed here that the fluid obeys an ideal gas equation of state, $p = \left(\gamma - 1\right) \rho e$, so that $h = \gamma e = C_p T$, where $C_p$ is the specific heat at constant pressure, $T$ is the temperature and $\gamma = C_p/C_v$ is the adiabatic index ($C_v$ is the specific heat at constant volume). The integration constants in equations (\ref{eq:veq}) and (\ref{eq:heq}) have been expressed in terms of both pre-shock (denoted by a subscript ``0'') and post-shock (denoted by a subscript ``1'') velocities using the shock compression ratio,
\be\label{eq:R}
R \equiv \frac{\rho_1}{\rho_0} = \frac{\gamma+1}{\gamma - 1 + 2/M_0^2},
\ee
where $M_0^2 = v_0^2/c_0^2$ is the shock Mach number and $c_0 = \sqrt{\gamma p_0/\rho_0}$ is the adiabatic sound speed in the ambient fluid \citep{Landau87}. The Prandtl number is given by $\Pran \equiv \mu C_p/\kappa$.

\section{Solutions}\label{sec:solutions}

\subsection{Becker's $\left(\Pran = 3/4\right)$ solution}\label{sec:pran34}

For $\Pran = 3/4$, equations (\ref{eq:veq}) and (\ref{eq:heq}) can be reduced to the quadrature \citep{Becker22}
\be\label{eq:integral_pran34}
x  = \frac{2L_\kappa }{\gamma + 1} \int \frac{\left(\kappa/\kappa_0\right)\eta}{\left(\eta - 1\right)\left(\eta - \eta_1\right)} \, d\eta,
\ee
and the algebraic expression
\be\label{eq:temp_pran34}
\frac{T}{T_0} = \frac{\gamma-1}{2}M_0^2\left(\frac{\gamma+1}{\gamma-1}  \eta_1 - \eta^2\right).
\ee
Here $\eta \equiv v/v_0$, $\eta_1 \equiv v_1/v_0 = R^{-1}$ and $L_\kappa \equiv \kappa_0/(m_0 C_v)$ is the ambient conductive length scale. For constant $\kappa = \kappa_0$, the integral (\ref{eq:integral_pran34}) is given by (to within an arbitrary constant)
\be\label{eq:solution_pran34}
x = \frac{2L_{\kappa}}{\gamma + 1} \ln \left[\left(1 - \eta\right)^\frac{1}{1 - \eta_1}\left(\eta - \eta_1\right)^{-\frac{\eta_1}{1 - \eta_1}}\right],
\ee
which is in turn equivalent to
\be\label{eq:solution2_pran34}
\left(\delta - \delta_1\right) f^{R-1} = \left(-\delta\right)^{R},
\ee
where $\delta \equiv \eta - 1$, $\delta_1 \equiv \eta_1 - 1 = R^{-1} - 1$,
\be\label{eq:f}
f \equiv e^{x/w},
\ee
and
\be\label{eq:length}
w \equiv \frac{2L}{\gamma+1}.
\ee
Here $L = L_\kappa$, but expression (\ref{eq:length}) is kept general for use in later sections.

\setlength{\tabcolsep}{10pt}
\begin{table}
  \begin{center}
\def~{\hphantom{0}}
  \begin{tabular}{ccc}
      $R$  & $M_0^2$   &   Equation \\[3pt]
       4/3   & $8/(7 - \gamma)$ & $\delta^4 - f\left(\delta ^3 + 3 \delta ^2/4 + 3\delta/16 + 1/64\right) = 0$ \\
       3/2  & $6/(5 - \gamma)$ & $\delta^3 + f\left(\delta ^2  + 2\delta/3 + 1/9\right) = 0$ \\
       2   & $4/(3-\gamma)$ & $\delta^2 - f\left(\delta + 1/2\right) = 0$ \\
       3 & $3/(2 - \gamma)$ & $\delta^3 + f^2\left(\delta + 2/3\right)  = 0$ \\
       4   & $8/(5-3\gamma)$ & $\delta^4 - f^3\left(\delta + 3/4\right) = 0$ \\
  \end{tabular}
  \caption{$\Pran = 3/4$ and $\Pran = \infty$ polynomials}
  \label{tab:pran34}
  \end{center}
\end{table}

\begin{figure}
\centering
\begin{tabular}{cc}
  \includegraphics[scale=0.4]{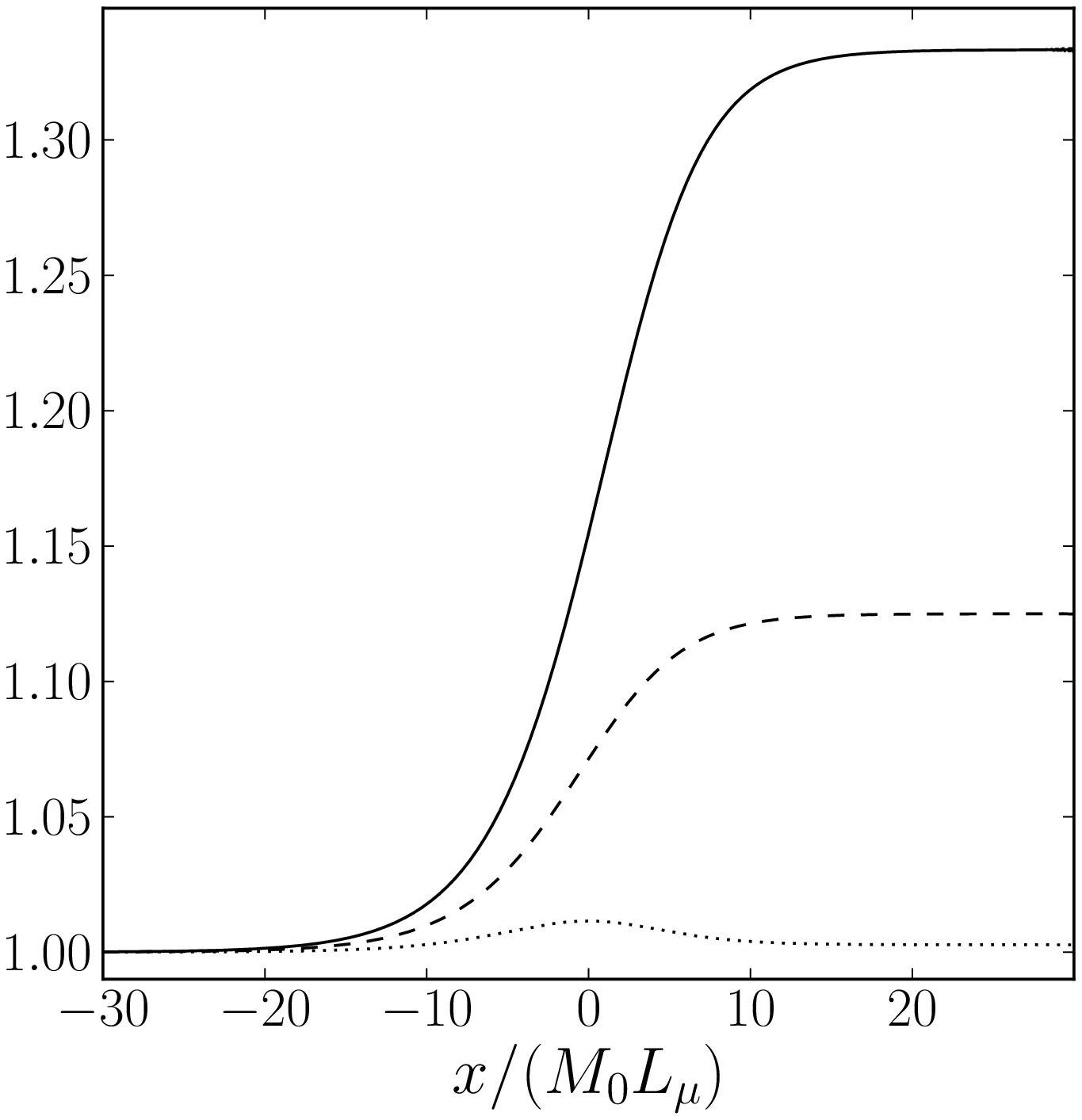} &
  \hspace{-0.0in}
  \includegraphics[scale=0.4]{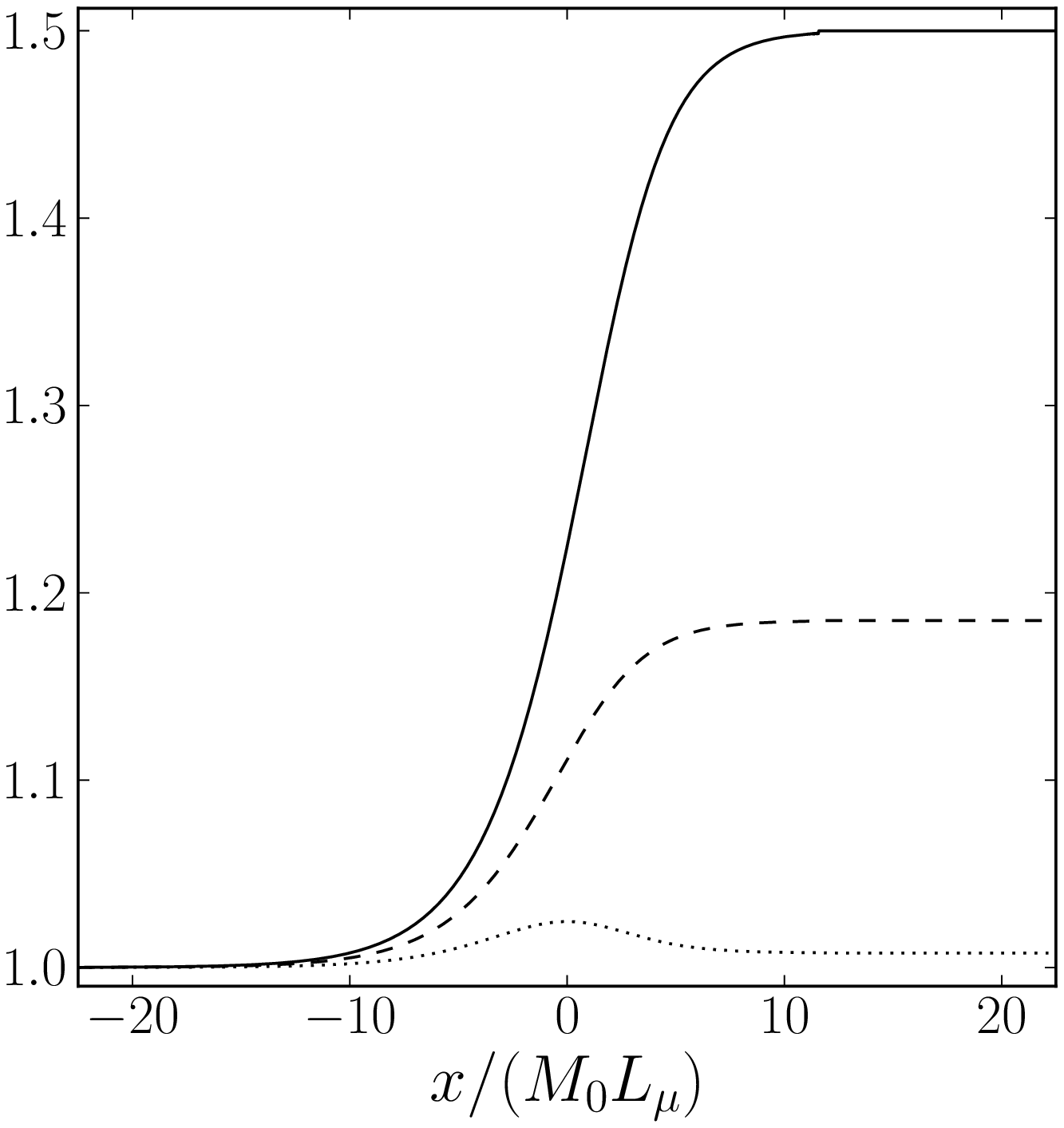}
\end{tabular}
  \caption{Curves of density (\emph{solid}), temperature (\emph{dashed}) and a proxy for the entropy (\emph{dotted}) for $\Pran = 3/4$ solutions with $R = 4/3$ (\emph{left}) and $R = 3/2$ (\emph{right}).}
\label{fig:R43_R32}
\end{figure}

\begin{figure}
\centering
\begin{tabular}{cc}
  \includegraphics[scale=0.4]{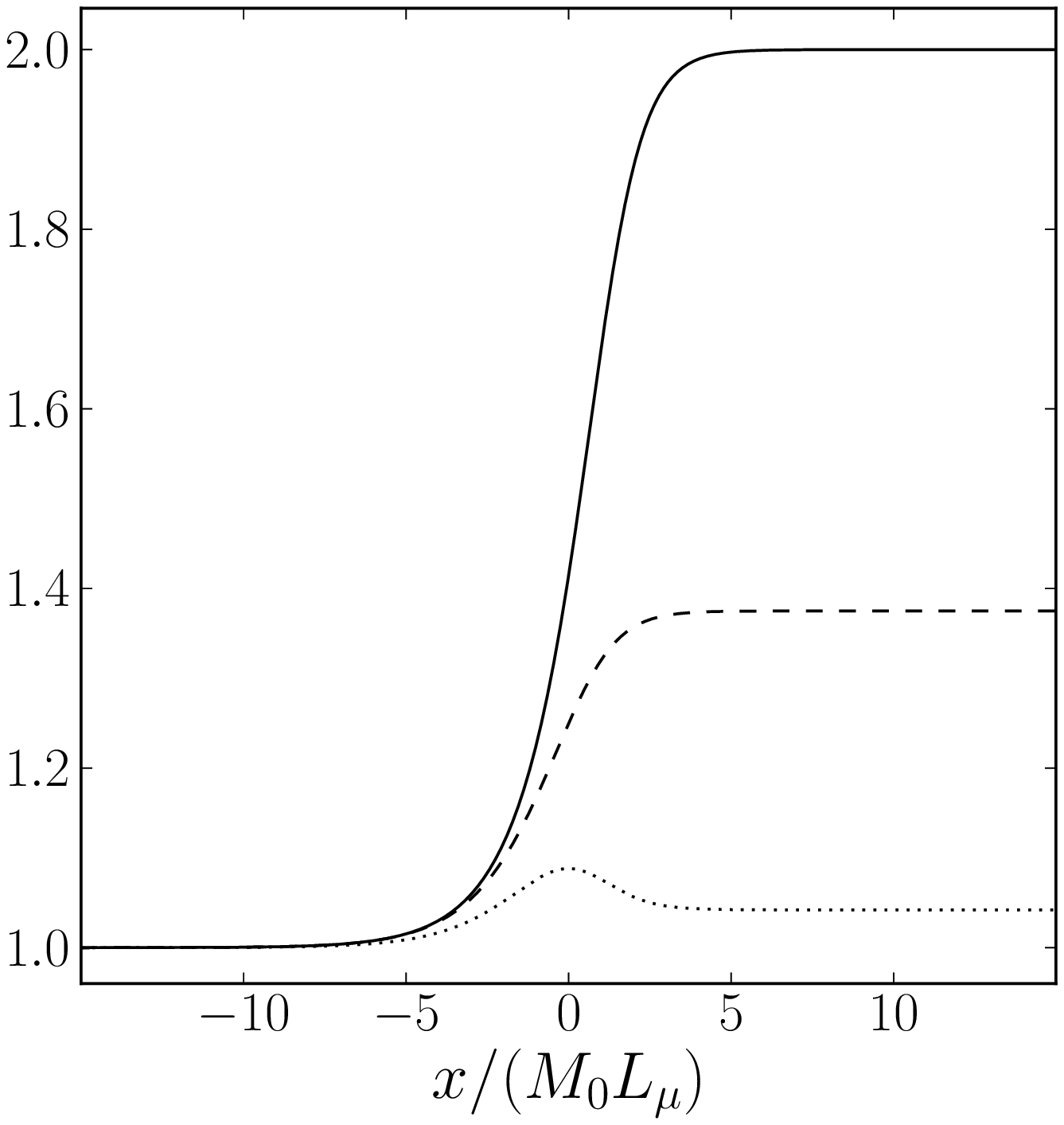} &
  \hspace{-0.0in}
  \includegraphics[scale=0.4]{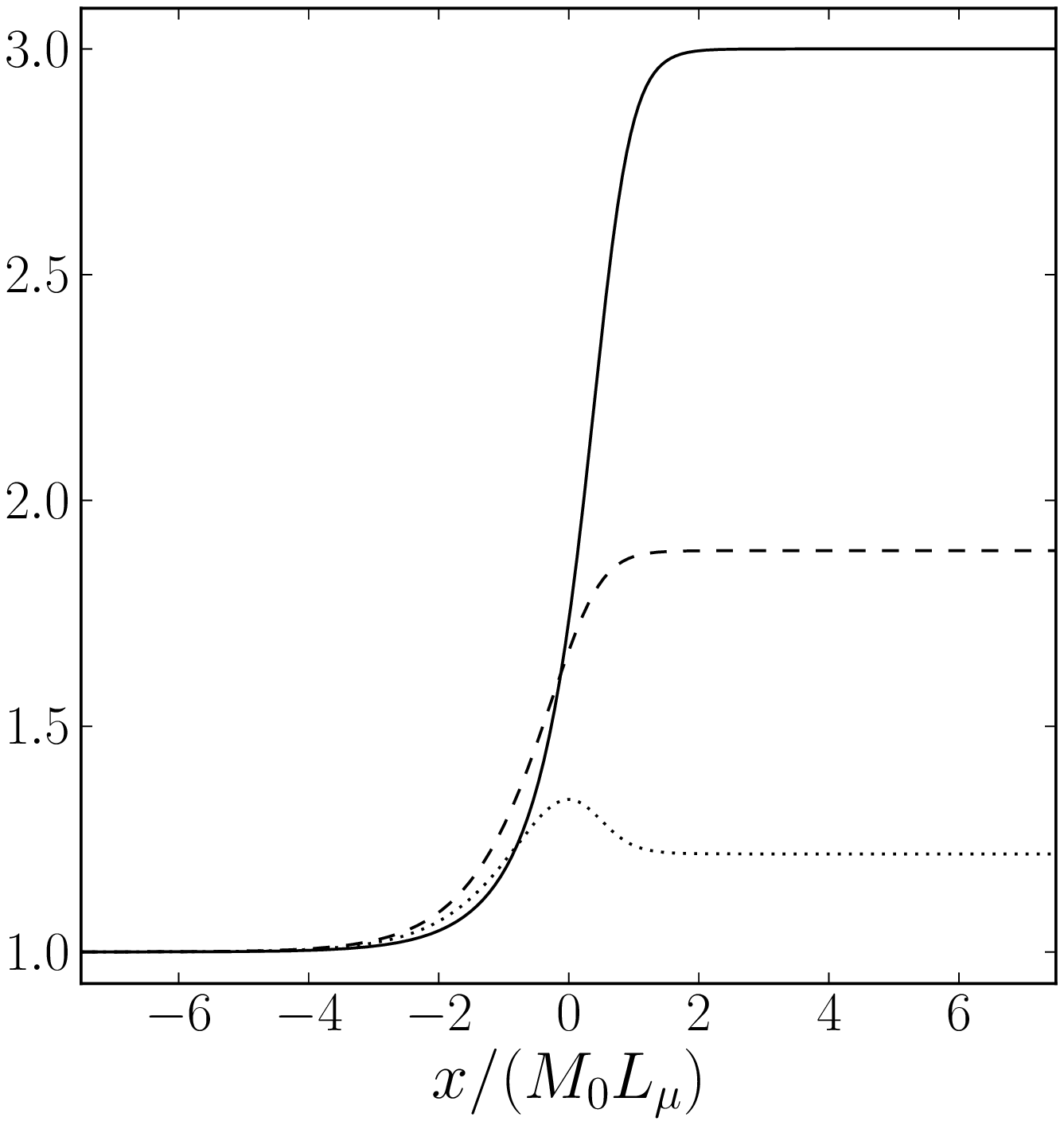}
\end{tabular}
  \caption{Curves of density (\emph{solid}), temperature (\emph{dashed}) and a proxy for the entropy (\emph{dotted}) for $\Pran = 3/4$ solutions with $R = 2$ (\emph{left}) and $R = 3$ (\emph{right}).}
\label{fig:R2_R3}
\end{figure}

\begin{figure}
\centering
\begin{tabular}{cc}
  \includegraphics[scale=0.4]{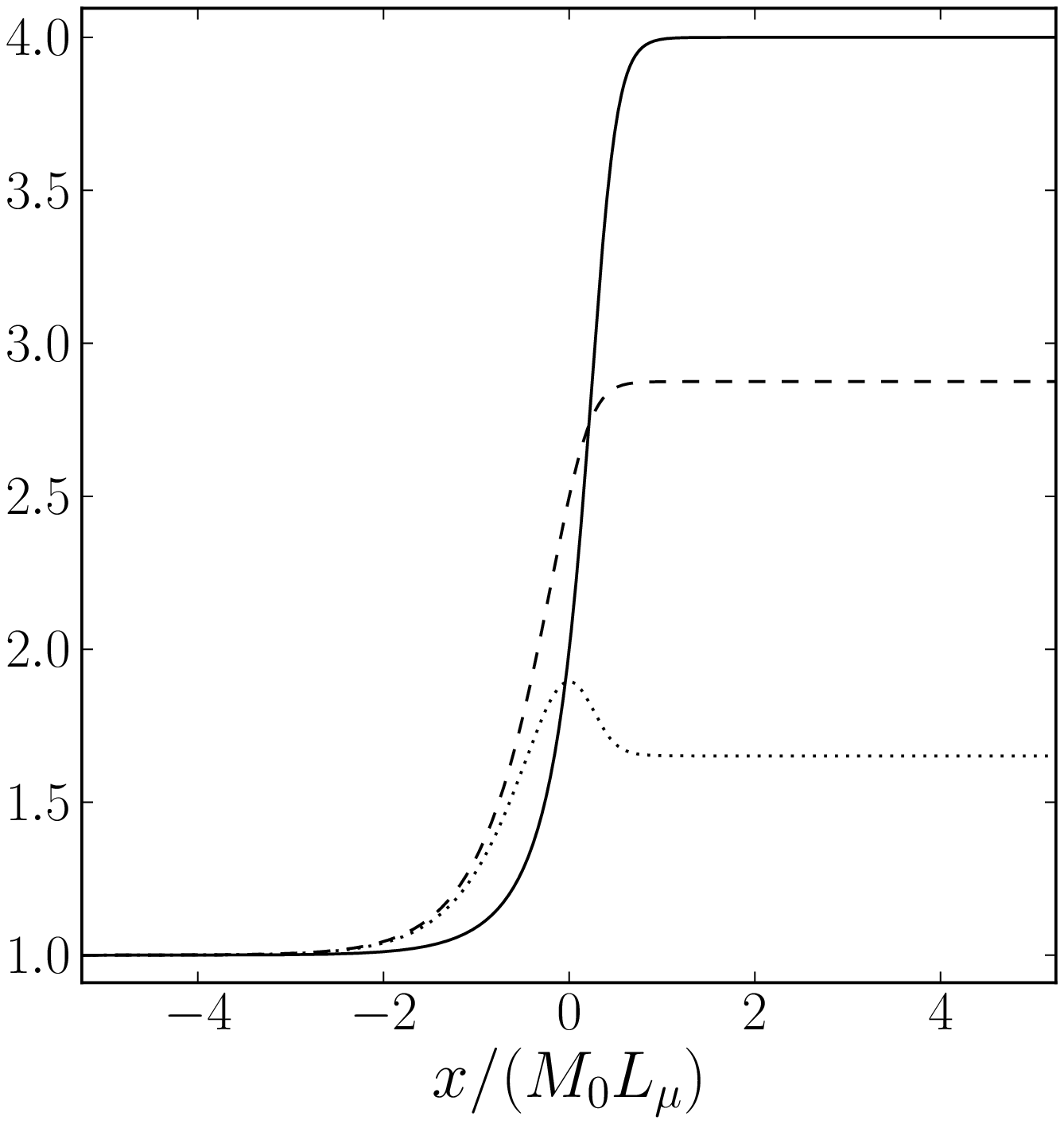} &
  \hspace{-0.0in}
  \includegraphics[scale=0.4]{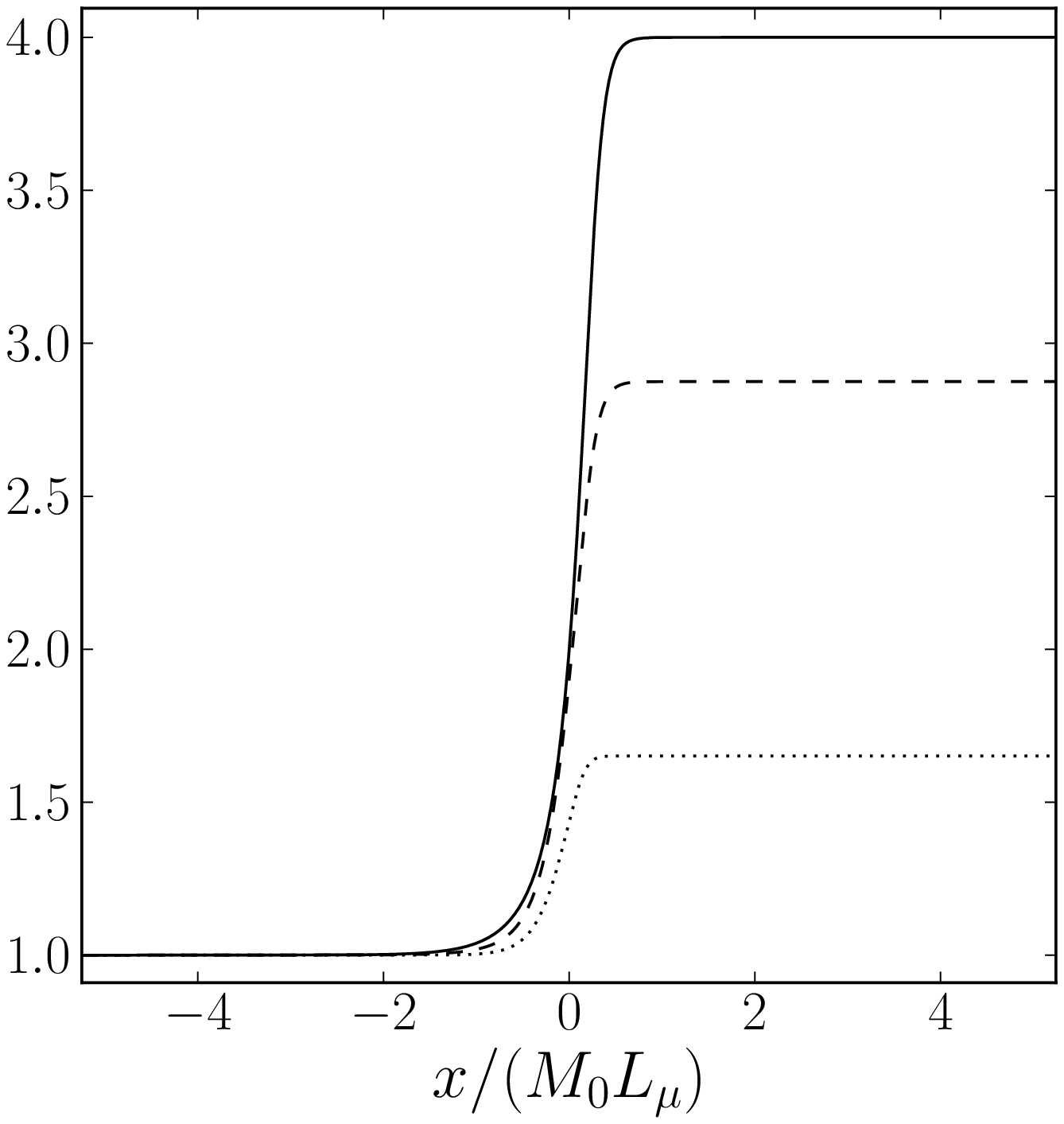}
\end{tabular}
  \caption{Curves of density (\emph{solid}), temperature (\emph{dashed}) and a proxy for the entropy (\emph{dotted}) for a $\Pran = 3/4$ solution (\emph{left}) and a $\Pran = \infty$ solution (\emph{right}) with $R = 4$.}
\label{fig:R4}
\end{figure}

For rational values of $R$, equation (\ref{eq:solution2_pran34}) is a polynomial in $\delta(x)$. Values for $R$ that yield closed-form expressions for $v(x)$ are listed in table~\ref{tab:pran34}, the corresponding closed-form expressions for $\eta$ are given in the appendix, and plots of the density, temperature and a proxy for the entropy ($s \equiv T \eta^{\gamma-1}$) are shown in figures~\ref{fig:R43_R32}--\ref{fig:R4}. Plotted quantities are all normalized to their ambient values, and the $x$ values have been scaled to $M_0 L_\mu$ ($= M_0 L_\kappa/\gamma$ for $\Pran = 3/4$), as this is a length scale that is independent of the shock Mach number. Table~\ref{tab:pran34} also gives the curves in $M_0$--$\gamma$ space for which the closed-form solutions are valid. These can be obtained by solving expression ($\ref{eq:R}$) for $M_0$:
\be
M_0 = \sqrt{\frac{2R}{R+1-\gamma(R-1)}}
\ee

\subsection{Large-$\Pran$ solution}\label{sec:praninf}

For $\Pran \rightarrow \infty$, equations (\ref{eq:veq}) and (\ref{eq:heq}) can be reduced to the quadrature \citep{Taylor10,Johnson13}
\be\label{eq:integral_praninf}
x  = \frac{2 L_\mu}{\gamma + 1} \int \frac{\left(\mu/\mu_0\right) \eta}{\left(\eta - 1\right)\left(\eta - \eta_1\right)} \, d\eta,
\ee
where $L_\mu \equiv 4\mu_0/(3 m_0)$ is the ambient viscous length scale, and the algebraic expression
\be\label{eq:temp_praninf}
T = T_0 \frac{\gamma (\gamma-1)M_0^2 }{2}\left(\eta^2 - 4\eta_i \eta + \frac{\gamma+1}{\gamma-1}  \eta_1\right),
\ee
where
\be\label{eq:etai}
\eta_i \equiv \frac{\gamma+1}{4\gamma}\left(1 + \eta_1\right).
\ee
For constant $\mu = \mu_0$, the integral (\ref{eq:integral_praninf}) is given by (to within an arbitrary constant)
\be\label{eq:solution_praninf}
x = \frac{2L_{\mu}}{\gamma + 1} \ln \left[\left(1 - \eta\right)^\frac{1}{1 - \eta_1}\left(\eta - \eta_1\right)^{-\frac{\eta_1}{1 - \eta_1}}\right].
\ee
Comparing expression (\ref{eq:solution_praninf}) with (\ref{eq:solution_pran34}), it can be seen that the solutions for $\eta$ in this limit are the same as those of the previous section with $L = L_\mu$ in expression (\ref{eq:length}). Figure~\ref{fig:R4} compares the large-$\Pran$ solution with $R = 4$ to the corresponding $\Pran = 3/4$ solution. Notice that the entropy has no local maximum in this limit (it increases monotonically from pre- to post-shock). This can be seen from
\be
\dv{\ln T}{x} + (\gamma-1)\dv{\ln \eta}{x} = 0 \rightarrow \eta^2 - \left(1 + \eta_1\right)\eta + \eta_1 = 0,
\ee
which is solved by $\eta = 1$ and $\eta = \eta_1$; the entropy has zero slope only at the boundaries.

\subsection{Small-$\Pran$ solution}\label{sec:pranzero}

For $\Pran \rightarrow 0$, equations (\ref{eq:veq}) and (\ref{eq:heq}) can be reduced to the quadrature \citep{Taylor10,Johnson13}
\be\label{eq:integral_pranzero}
x =  \frac{4 L_\kappa}{\gamma+1} \int \frac{\left(\kappa/\kappa_0\right)\left(\eta  - \eta_i\right)}{\left(\eta - 1\right)\left(\eta - \eta_1\right)}\,d\eta,
\ee
and the algebraic expression
\be\label{eq:temp_pranzero}
T = T_0 \gamma M_0^2 \eta\left(2\eta_i  - \eta\right).
\ee
For constant $\kappa = \kappa_0$, the integral (\ref{eq:integral_pranzero}) is given by (to within an arbitrary constant)
\be\label{eq:solution_pranzero}
x =  \frac{4L_\kappa}{\gamma+1} \ln \left[\left(1 - \eta\right)^{\frac{1-\eta_i}{1 - \eta_1}}\left(\eta - \eta_1\right)^{\frac{\eta_i - \eta_1}{1 - \eta_1}}\right],
\ee
which is in turn equivalent to
\be\label{eq:solution2_pranzero}
f^{\left|n-1\right|/2}\left(\delta - \delta_1\right) = \left(-\delta\right)^n,
\ee
where
\be\label{eq:n}
n \equiv \frac{\eta_i-1}{\eta_i - \eta_1} = \frac{(\gamma+1)(1-\gamma M_0^2)}{1-3\gamma+(3-\gamma)\gamma M_0^2},
\ee
$f$ is defined in expression (\ref{eq:f}), and $L = L_\kappa$ in expression (\ref{eq:length}). For $M_0^2 > 1$, one has $1 < |n| < \infty$.

\begin{figure}
\centering
\begin{tabular}{cc}
  \includegraphics[scale=0.4]{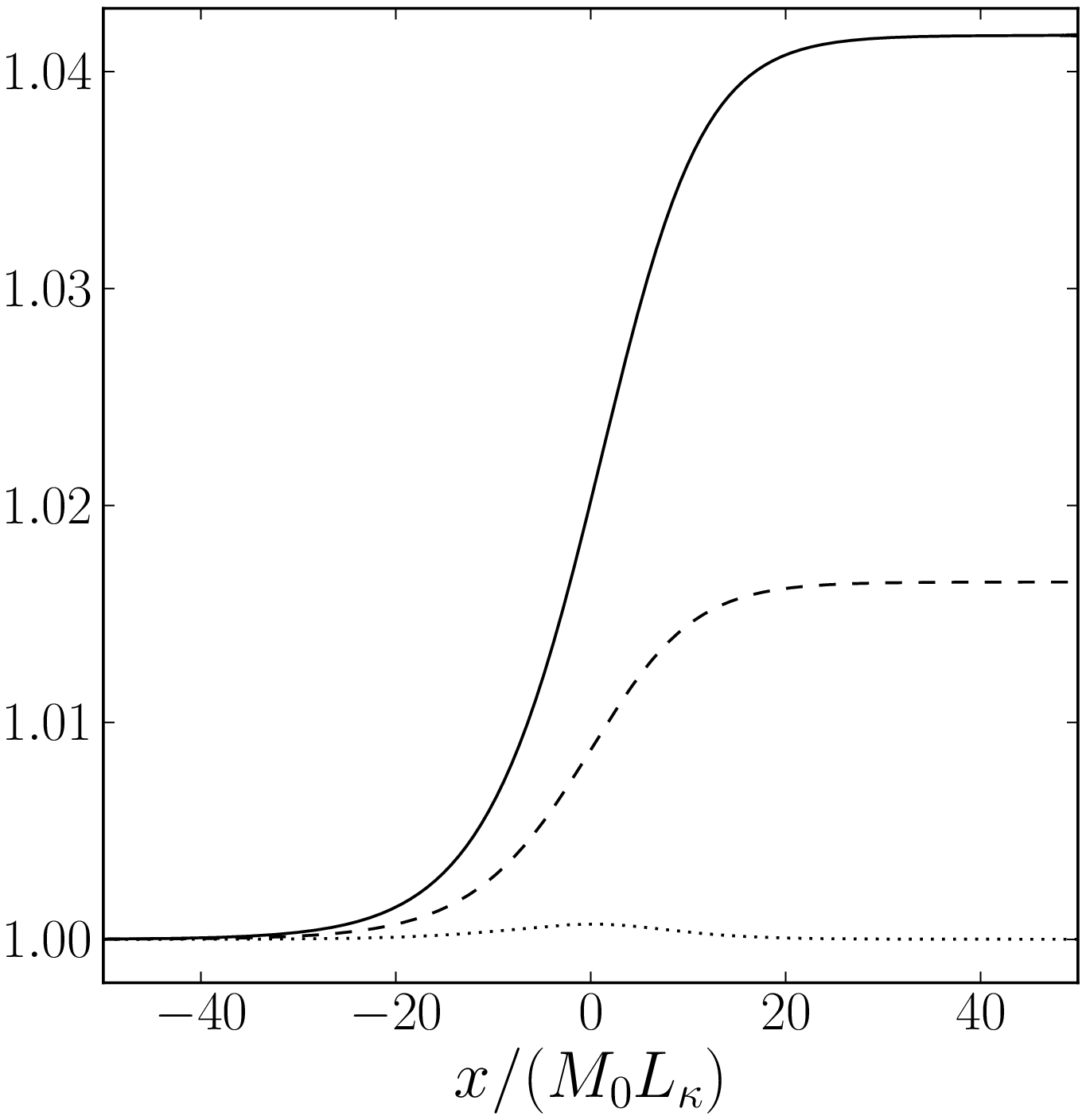} &
  \hspace{-0.0in}
  \includegraphics[scale=0.4]{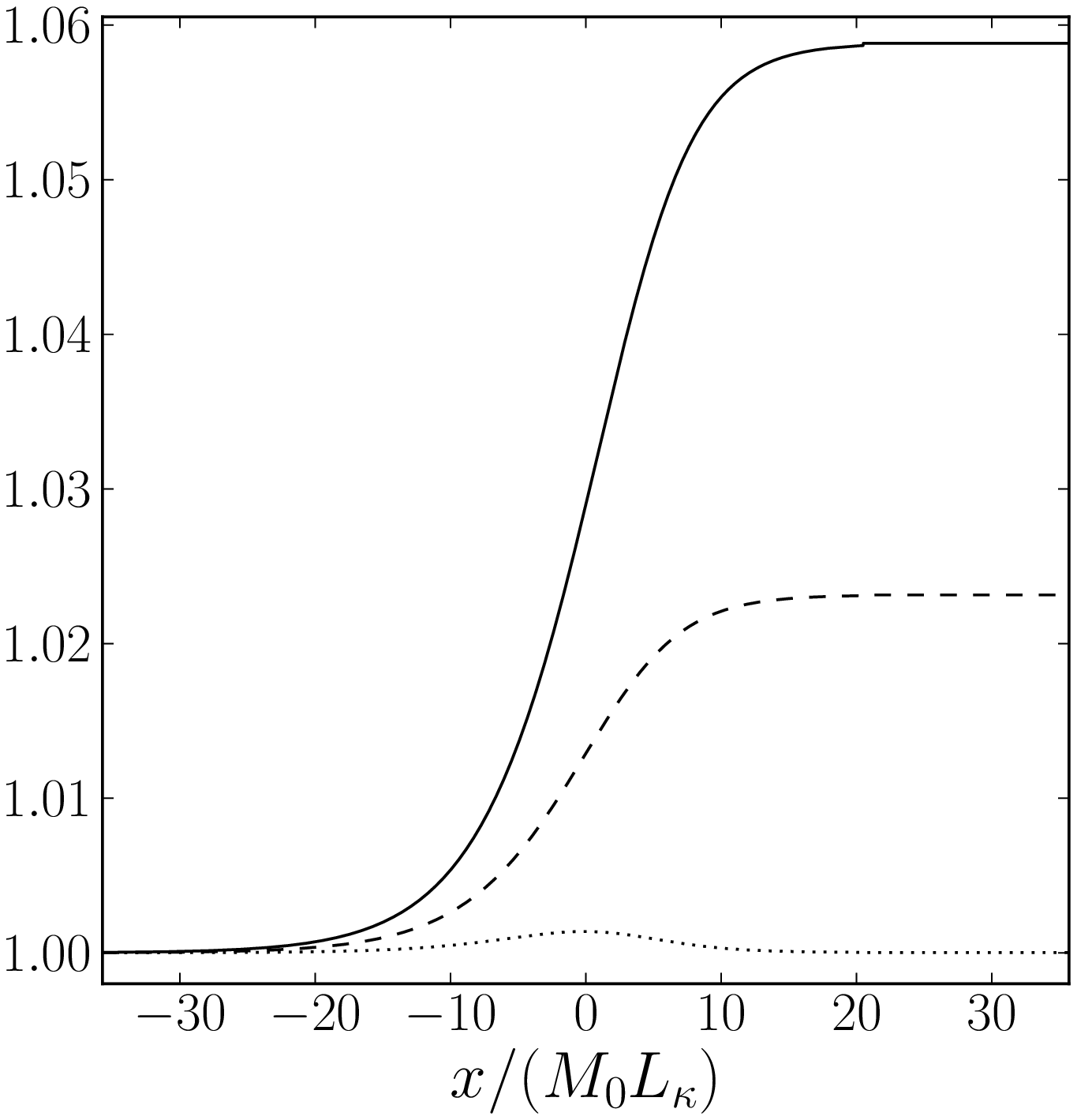}
\end{tabular}
  \caption{Curves of density (\emph{solid}), temperature (\emph{dashed}) and a proxy for the entropy (\emph{dotted}) for $\Pran = 0$ solutions with $n = 4/3$ (\emph{left}) and $n = 3/2$ (\emph{right}).}
\label{fig:n43_n32}
\end{figure}

\begin{figure}
\centering
\begin{tabular}{cc}
  \includegraphics[scale=0.4]{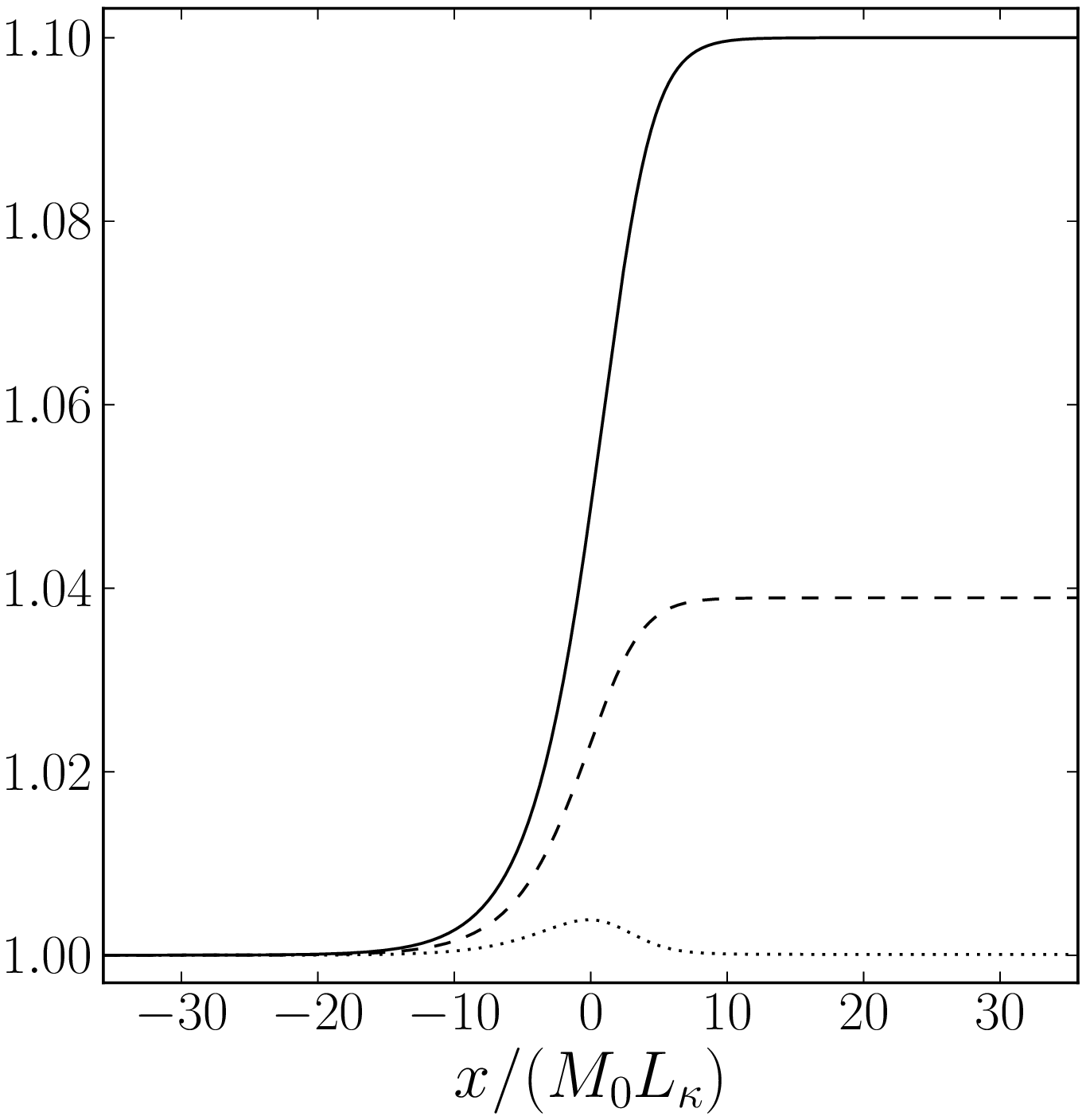} &
  \hspace{-0.0in}
  \includegraphics[scale=0.4]{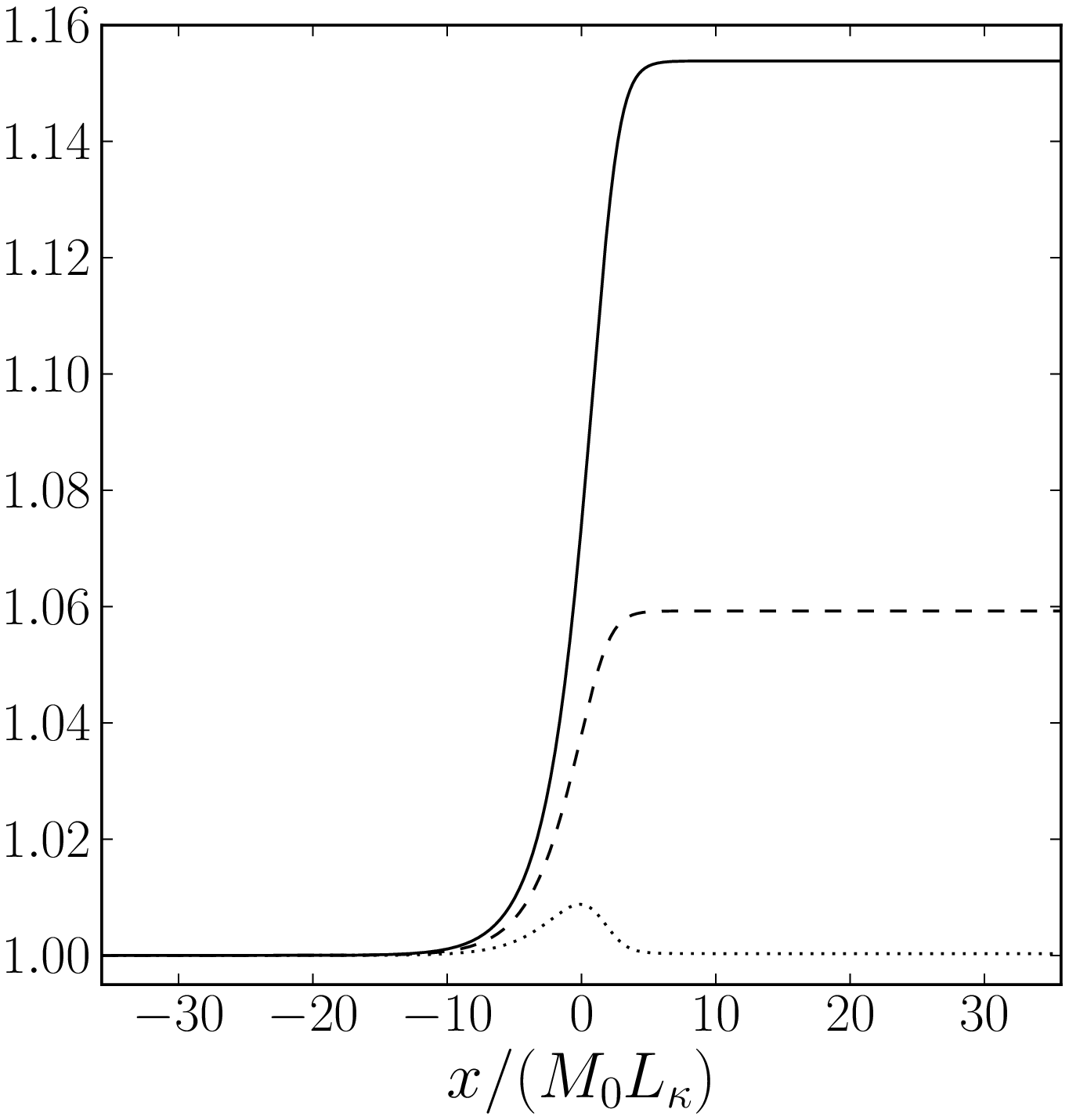}
\end{tabular}
  \caption{Curves of density (\emph{solid}), temperature (\emph{dashed}) and a proxy for the entropy (\emph{dotted}) for $\Pran = 0$ solutions with $n = 2$ (\emph{left}) and $n = 3$ (\emph{right}).}
\label{fig:n2_n3}
\end{figure}

\begin{figure}
\centering
\begin{tabular}{cc}
  \includegraphics[scale=0.4]{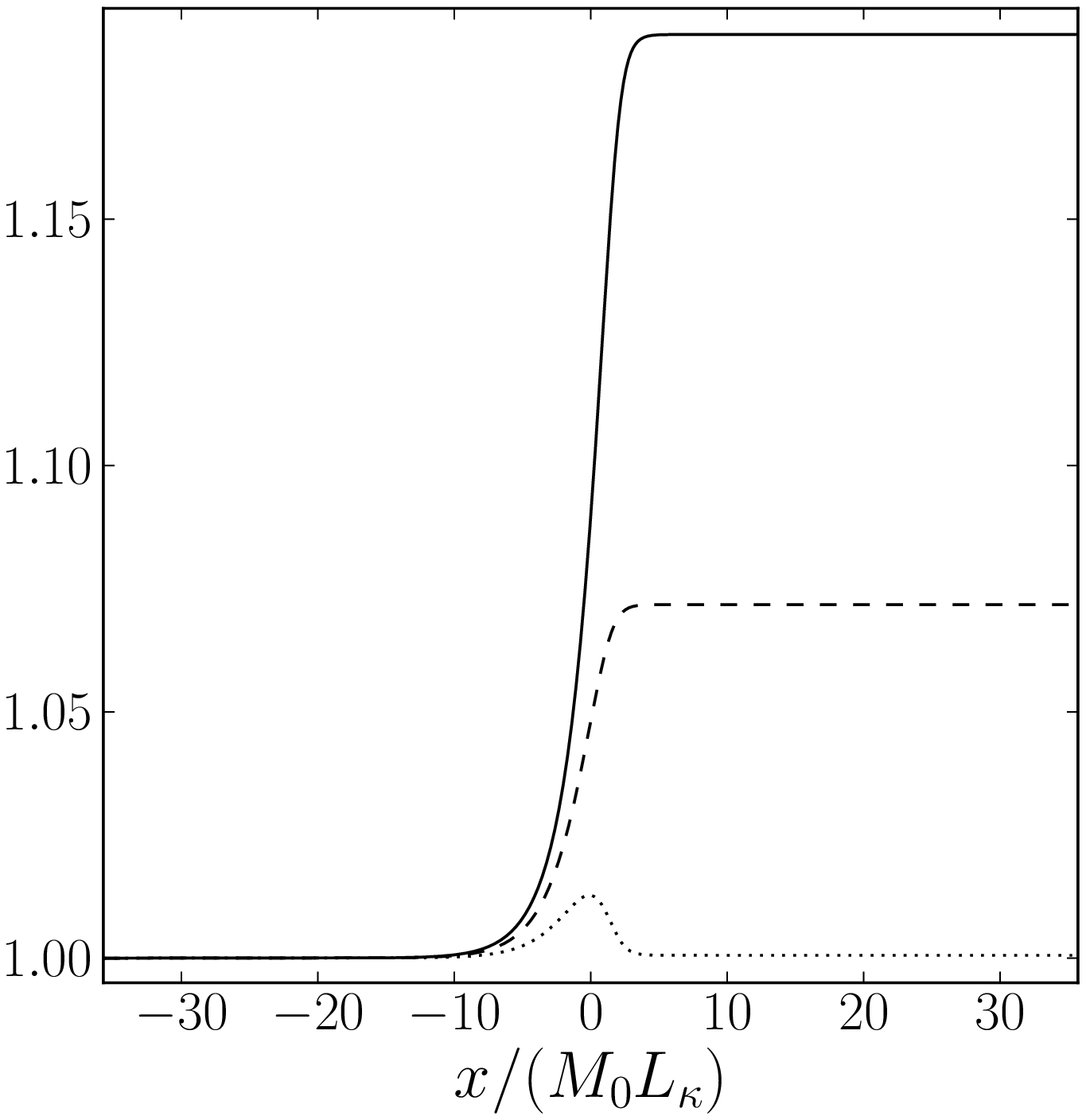} &
  \hspace{-0.0in}
  \includegraphics[scale=0.4]{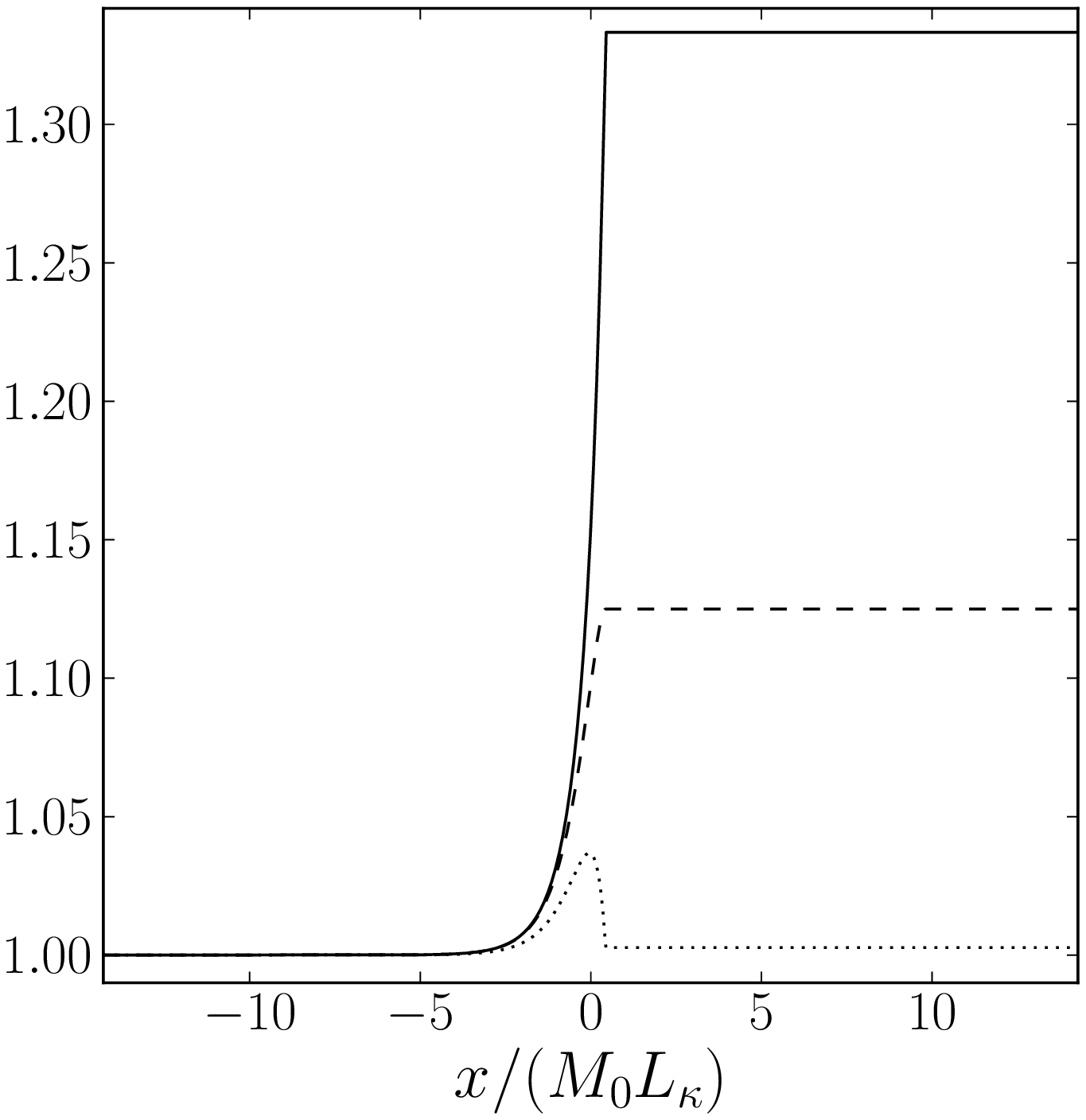}
\end{tabular}
  \caption{Curves of density (\emph{solid}), temperature (\emph{dashed}) and a proxy for the entropy (\emph{dotted}) for $\Pran = 0$ solutions with $n = 4$ (\emph{left}) and $n = \infty$ (\emph{right}).}
\label{fig:n4_ninf}
\end{figure}

\begin{figure}
\centering
\begin{tabular}{cc}
  \includegraphics[scale=0.4]{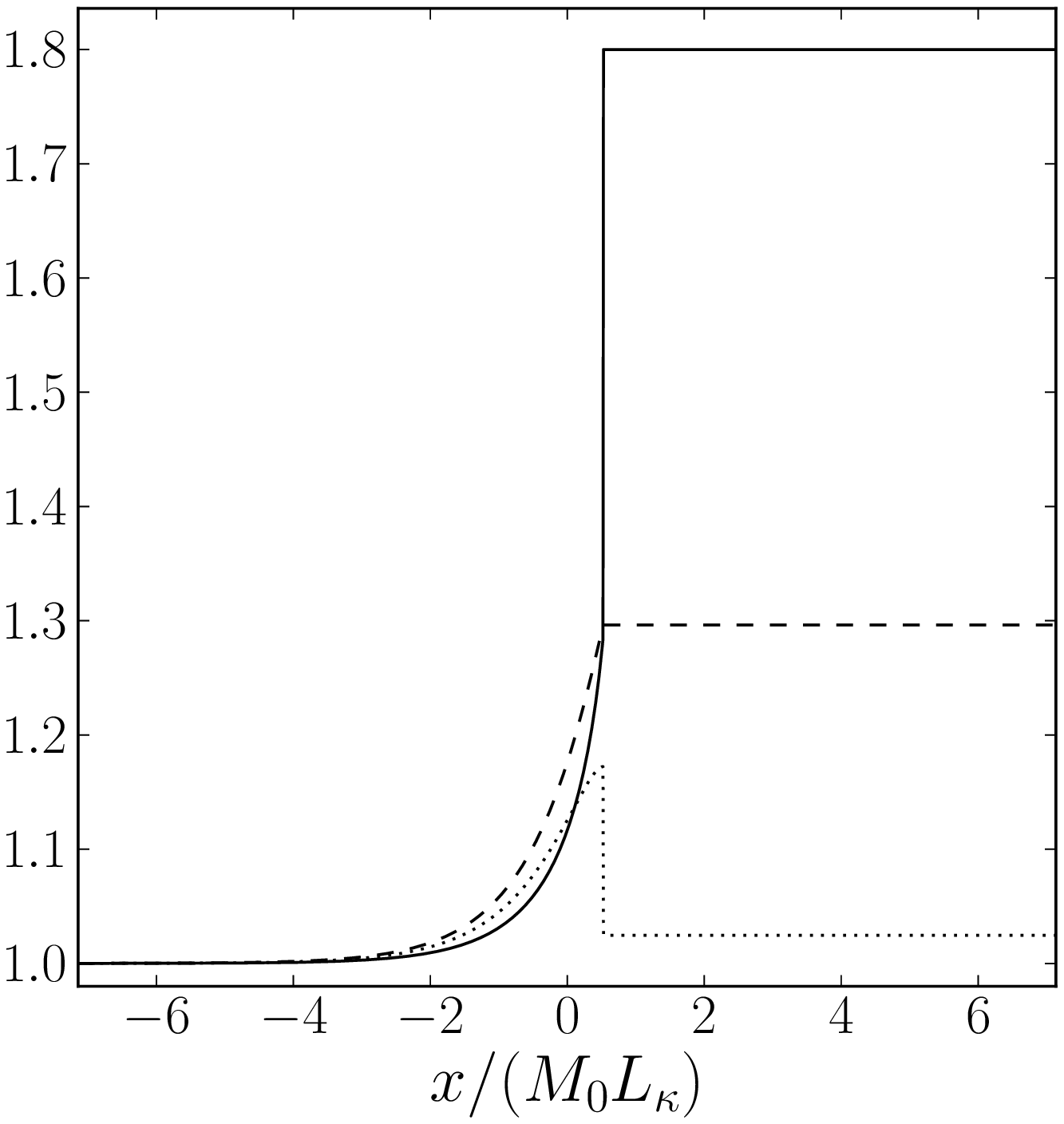} &
  \hspace{-0.0in}
  \includegraphics[scale=0.4]{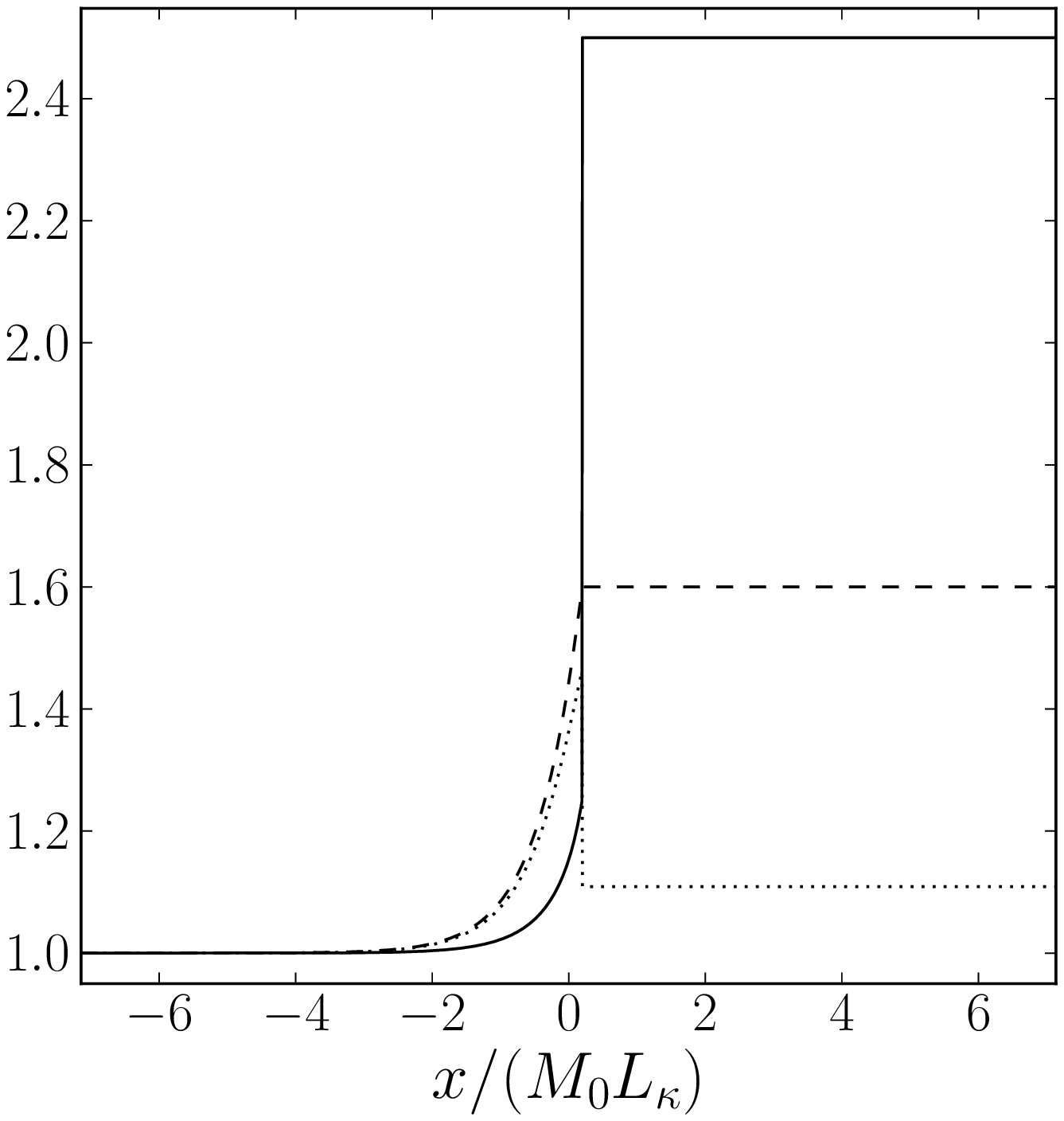}
\end{tabular}
  \caption{Curves of density (\emph{solid}), temperature (\emph{dashed}) and a proxy for the entropy (\emph{dotted}) for $\Pran = 0$ solutions with $n = -3$ (\emph{left}) and $n = -2$ (\emph{right}).}
\label{fig:nm3_nm2}
\end{figure}

For rational values of $n$, equation (\ref{eq:solution2_pranzero}) is a polynomial in $\delta(x)$. Values for $n$ that yield closed-form expressions for $v(x)$ are listed in table~\ref{tab:pranzero}, the corresponding closed-form expressions for $\eta$ are given in the appendix, and plots of the density, temperature and a proxy for the entropy are shown in figures~\ref{fig:n43_n32}--\ref{fig:nm3_nm2}. Plotted quantities are again normalized to their ambient values, and the $x$ values have been scaled to $M_0 L_\kappa$. Table~\ref{tab:pranzero} also gives the curves in $M_0$--$\gamma$ space for which the closed-form solutions are valid. These can be obtained by solving expression ($\ref{eq:n}$) for $M_0$:
\be
M_0 = \sqrt{\frac{\alpha\gamma-1}{\gamma(\alpha - \gamma)}},\;\;\alpha \equiv \frac{3n+1}{n-1}.
\ee
In terms of $\alpha$,
\be
\delta_1 = -2\frac{\gamma-1}{\alpha \gamma - 1}.
\ee

\setlength{\tabcolsep}{10pt}
\begin{table}
  \begin{center}
\def~{\hphantom{0}}
  \begin{tabular}{cccc}
      $n$ & $\alpha$ & $M_0^2$ & Equation \\[3pt]
      -3  & $2$ & $(2\gamma-1)/(\gamma[2 - \gamma])$ & $\delta^4 - \delta_1\delta^3 + f^2  = 0$ \\
       -2  & $5/3$ & $(5\gamma-3)/(\gamma[5 - 3\gamma])$ & $\delta^3 - \delta_1\delta^2 - f^{3/2}  = 0$ \\
       4/3 & $15$ & $(15\gamma-1)/(\gamma[15 - \gamma])$& $\delta^4 + f^{1/2}\left(-\delta^3 + 3\delta_1\delta^2 - 3\delta_1^2\delta + \delta_1^3\right)= 0$ \\
       3/2 & $11$ & $(11\gamma-1)/(\gamma[11 - \gamma])$ & $\delta^3 + f^{1/2}\left(\delta^2 - 2\delta_1 \delta + \delta_1^2\right) = 0$ \\
       2  & $7$ &$(7\gamma-1)/(\gamma[7 - \gamma])$ & $\delta^2 + f^{1/2}\left(-\delta + \delta_1\right) = 0$ \\
       3 & $5$ &$(5\gamma-1)/(\gamma[5 - \gamma])$ &  $\delta^3 + f\left(\delta - \delta_1\right)= 0$ \\
       4  & $13/3$ & $(13\gamma-3)/(\gamma[13 - 3\gamma])$&  $\delta^4 + f^{3/2}\left(-\delta + \delta_1\right) = 0$ \\
       $\infty$ & $3$ & $(3\gamma-1)/(\gamma[3 - \gamma])$ & $\delta + f^{1/2} = 0$ \\
  \end{tabular}
  \caption{$\Pran = 0$ polynomials}
  \label{tab:pranzero}
  \end{center}
\end{table}

For $M_0 > M_c$, where
\be
M_c \equiv \sqrt{\frac{3\gamma-1}{\gamma\left(3 - \gamma\right)}}
\ee
(this is equivalent to $n < 0$), the solution in this limit is discontinuous \citep{Zel'dovich02,Johnson13}. For $M_0 = M_c$, $n = \pm \infty$, $\eta_i = \eta_1$ and equation (\ref{eq:solution2_pranzero}) reduces to $f^{1/2} = -\delta$, or $\eta = 1 - f^{1/2}$. This solution is valid until $\eta = \eta_1$, where there is a weak discontinuity in both velocity and temperature. The weak discontinuity in the temperature occurs above the first derivative, since $dT/dx \propto \eta_i - \eta = 0$ at $\eta = \eta_i$.

\subsection{Constant kinematic viscosity}\label{sec:ckv}

\begin{figure}
\centering
\begin{tabular}{cc}
  \includegraphics[scale=0.4]{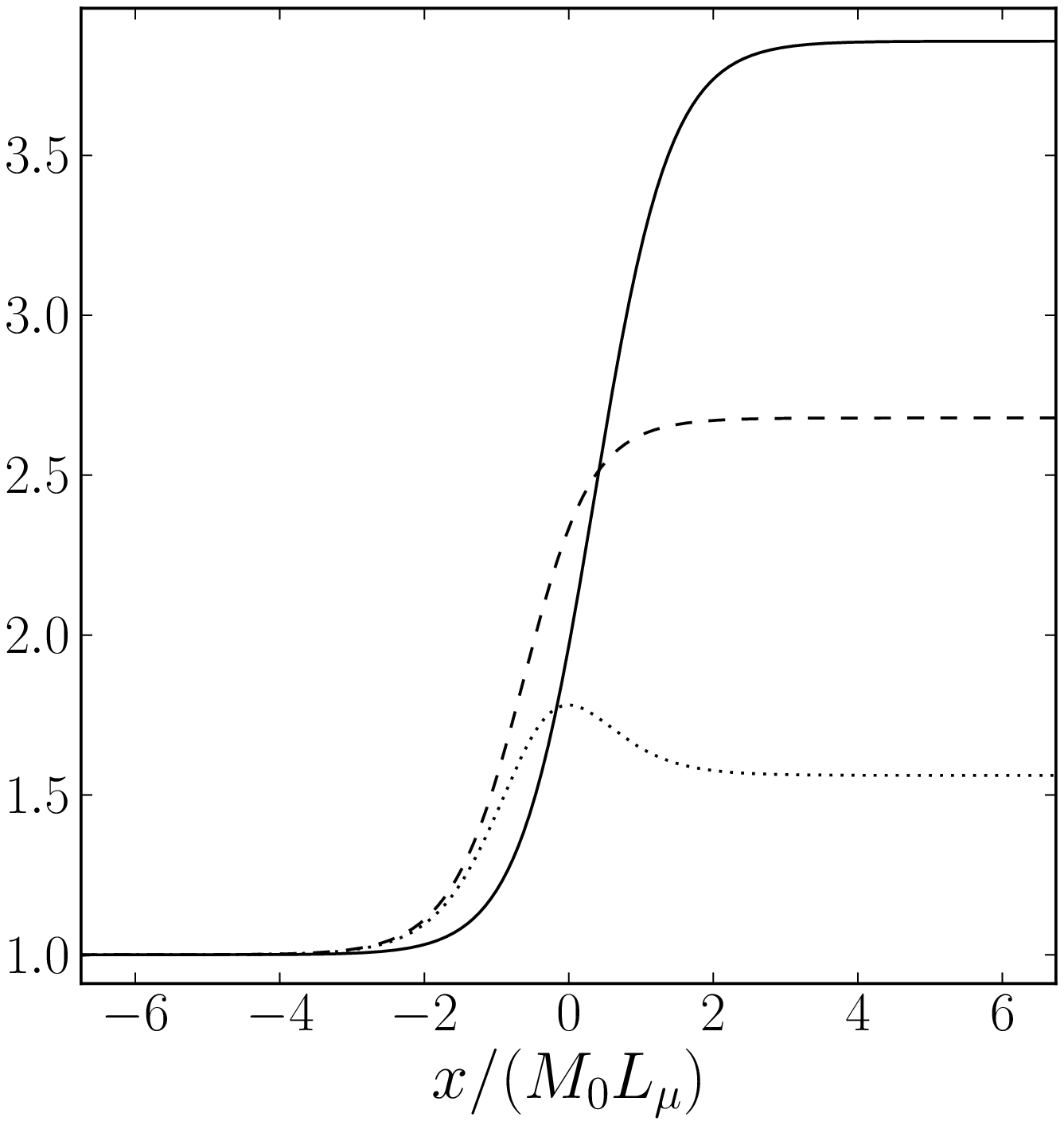} &
  \hspace{-0.0in}
  \includegraphics[scale=0.4]{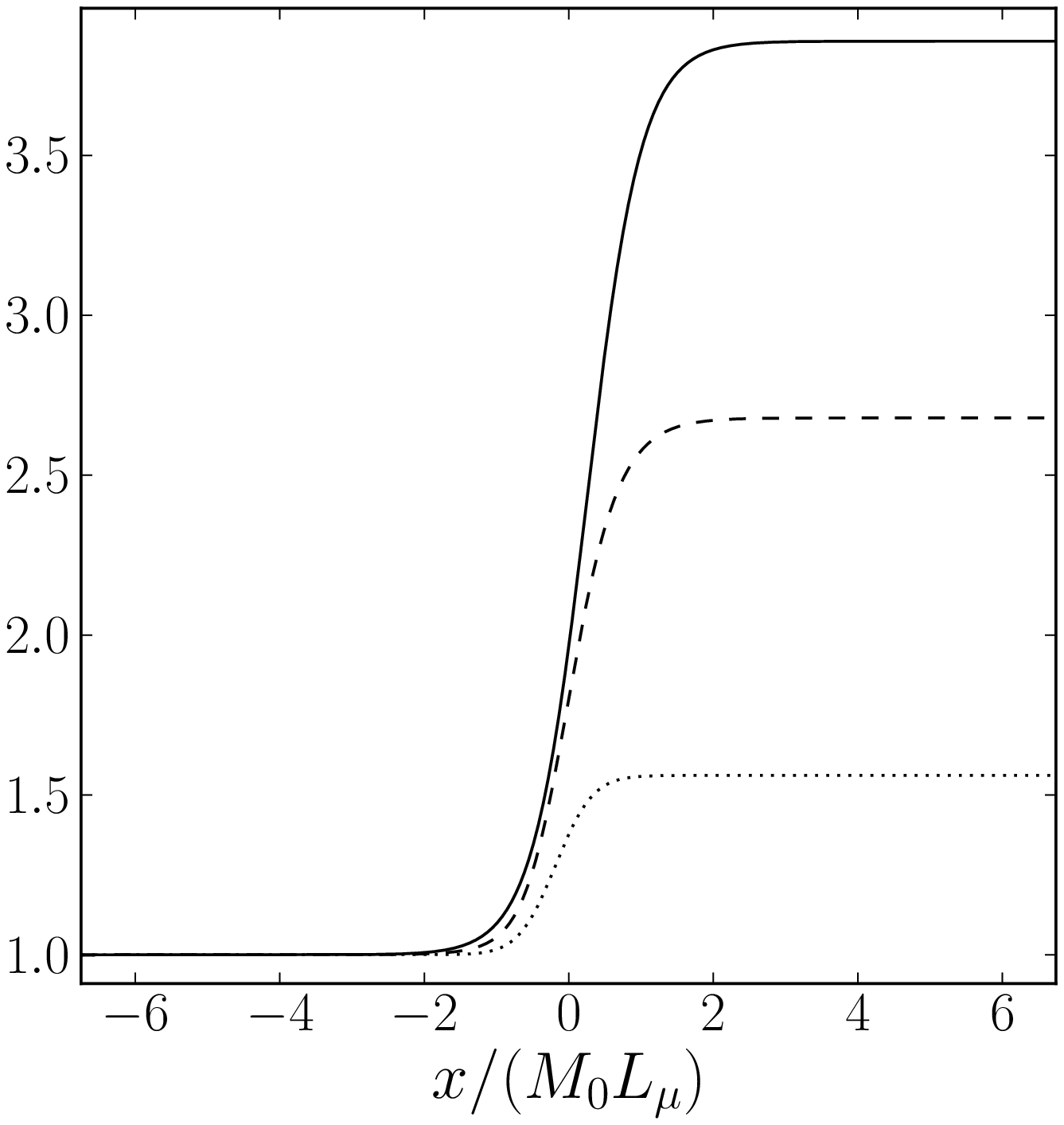}
\end{tabular}
  \caption{Curves of density (\emph{solid}), temperature (\emph{dashed}) and a proxy for the entropy (\emph{dotted}) for a $\Pran = 3/4$ solution (\emph{left}) and a $\Pran = \infty$ solution (\emph{right}) with constant $\nu$ and $M_0 = 3$.}
\label{fig:nu_const}
\end{figure}

For a constant kinematic viscosity, $\nu \equiv \mu/\rho = \nu_0$, the integrals (\ref{eq:integral_pran34}) and (\ref{eq:integral_praninf}) both reduce to
\be\label{eq:integral_nu0}
x = w \int \frac{d\eta}{\left(\eta - 1\right)\left(\eta - \eta_1\right)} = \frac{w}{1-\eta_1} \ln \left(\frac{1 - \eta}{\eta - \eta_1}\right),
\ee
which can be solved for $\eta$ to give
\be
\eta = \sigma\left(-z\right) + \eta_1 \, \sigma\left(z\right),
\ee
where
$$
\sigma(z) \equiv \frac{1}{1+e^{-z}},\;\;z \equiv \frac{x}{w}\left(1-\eta_1\right) = \frac{x}{L}\left(1-M_0^{-2}\right).
$$
This solution has the same form as the \cite{Taylor10} structure function for weak shocks; the latter was derived under the assumption of constant $\mu$ and $\kappa$. An equivalent expression for $\eta$ is
\be
\eta = \frac{\eta_1 + 1}{2} + \frac{\eta_1 - 1}{2} \tanh \left(\frac{z}{2}\right).
\ee
This solution is valid for both $\Pran = 3/4$, in which case $L = L_\kappa = \gamma L_\mu$ and $T$ is given by expression (\ref{eq:temp_pran34}), and $\Pran = \infty$, in which case $L = L_\mu$ and $T$ is given by expression (\ref{eq:temp_praninf}). Plots of the density, temperature and a proxy for the entropy (normalized to their ambient values) are shown in figure~\ref{fig:nu_const} for both $\Pran = 3/4$ and $\Pran = \infty$.

\section{Summary}\label{sec:summary}

\begin{figure}
\centering
\begin{tabular}{cc}
  \includegraphics[scale=0.4]{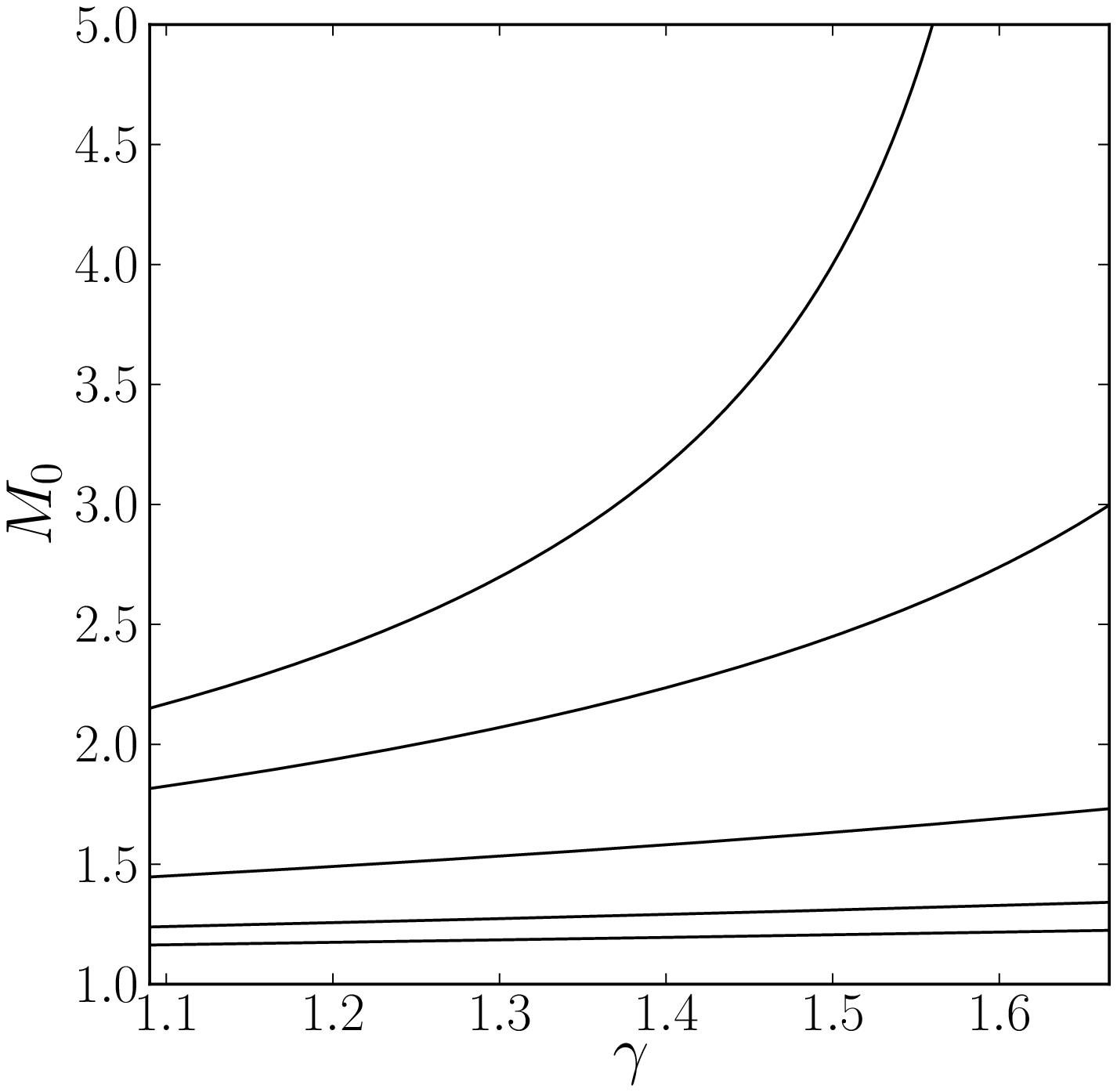} &
  \hspace{-0.0in}
  \includegraphics[scale=0.4]{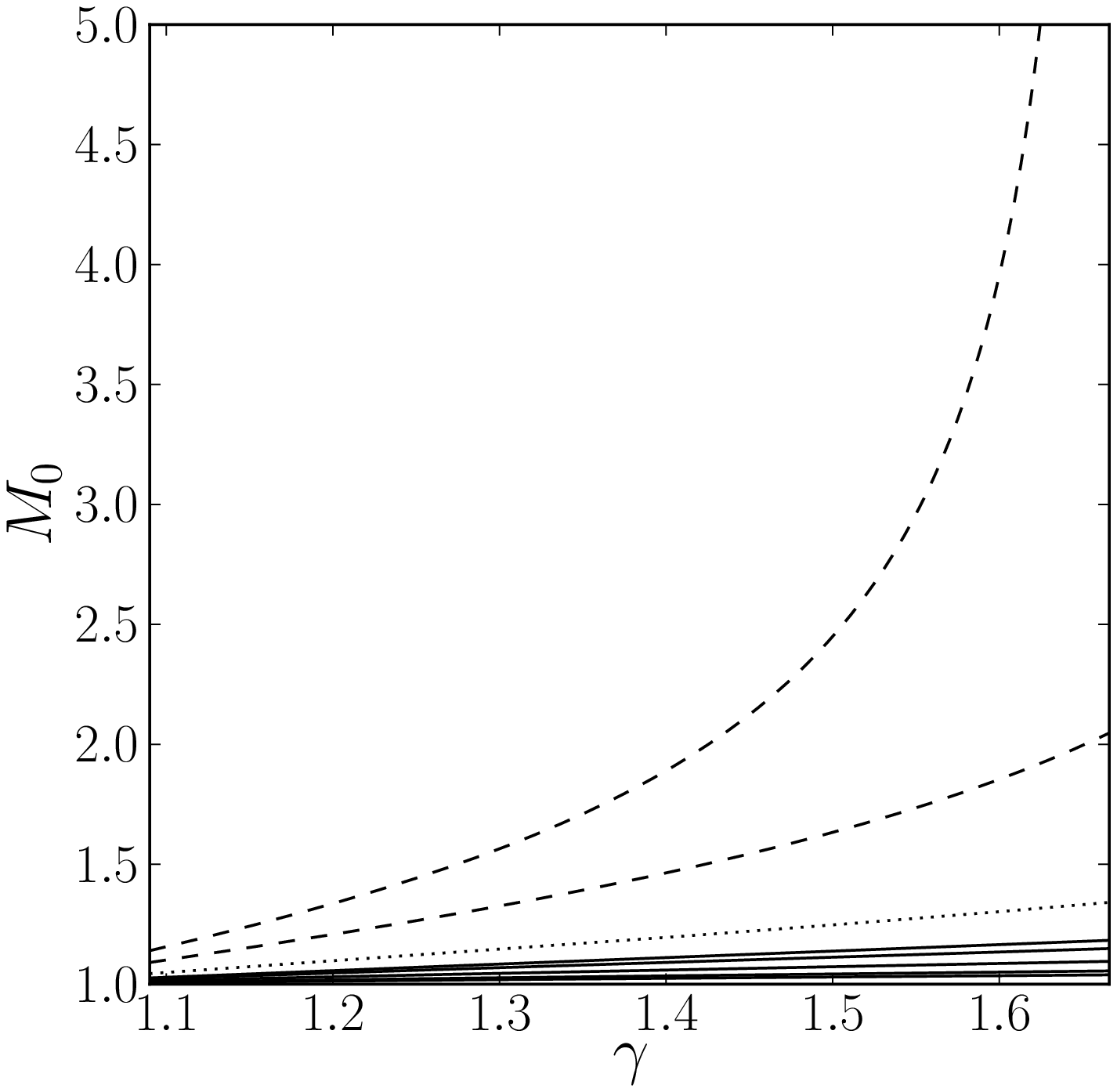}
\end{tabular}
  \caption{Curves in $M_0$--$\gamma$ space for which the derived closed-form solutions are valid, for $R = 4/3$, $3/2$, $2$, $3$ and $4$ (\emph{left, bottom to top}), and for $n = -3$, $-2$, $\infty$, $4/3$, $3/2$, $2$, $3$ and $4$ (\emph{right, top to bottom}). On the right, a dashed line indicates a discontinuous solution, a dotted line indicates a solution with a weak discontinuity, and a solid line indicates a continuous solution.}
\label{fig:M0_gamma}
\end{figure}

Several closed-form analytical solutions to the one-dimensional compressible Navier-Stokes equations have been derived in the limit of a steady state and an ideal gas equation of state. Solutions with a constant dynamic viscosity and thermal conductivity can be obtained by solving a polynomial equation. Polynomial solutions valid for large $\Pran$ and $\Pran = 3/4$ are listed in table~\ref{tab:pran34} and shown in figures~\ref{fig:R43_R32}--\ref{fig:R4}. Polynomial solutions valid for small $\Pran$ are listed in table~\ref{tab:pranzero} and shown in figures~\ref{fig:n43_n32}--\ref{fig:nm3_nm2}. Tables~\ref{tab:pran34} and \ref{tab:pranzero} also give expressions for $M_0(\gamma)$ for which these solutions are valid, and the corresponding curves in $M_0$--$\gamma$ space are shown in figure~\ref{fig:M0_gamma}. A solution can also be obtained under the assumption of a constant kinematic viscosity, valid for either large $\Pran$ or a constant $\Pran = 3/4$ and at any Mach number; this solution is described in \S\ref{sec:ckv} and shown in figure~\ref{fig:nu_const}.

The derived solutions are non-linear and exact in the sense that no source terms in the evolution equations are neglected or approximated. As such, they make excellent verification tests for numerical algorithms. The most physically relevant solutions are those with $\Pran = 3/4$, as this is close to the $\Pran$ of many gases. The small-$\Pran$ solutions are somewhat relevant to gas mixtures and plasmas, whereas the large-$\Pran$ solutions are primarily of academic interest and are only included for completeness \citep{Johnson13}. The derived solution set is not exhaustive: additional polynomial solutions exist under the assumption of a constant thermal diffusivity $\chi \equiv \kappa/\rho$, and a solution in terms of Lambert functions can be derived for $\mu \propto T^{1/2}$, $\Pran \rightarrow \infty$ and $M_0 \rightarrow \infty$. As none of these solutions are more physically relevant than the ones discussed above, their detailed derivation has not been included.

Perhaps the primary benefit of the derived solutions is their addition to the limited number of known exact solutions to the Navier-Stokes equations. Further study of the solutions may provide insight into the non-linear character of these equations, and the methods employed may stimulate additional analytical developments. 

I thank the referees for their comments. This work was performed under the auspices of Lawrence Livermore National Security, LLC, (LLNS), under Contract No.$\;$DE-AC52-07NA27344.

\appendix
\section{}\label{appA}

For the quadratic equations in tables~\ref{tab:pran34} and \ref{tab:pranzero} ($\delta^2 + a\delta + b = 0$), the solution branch relevant to a shock (the other solution branch grows exponentially as $x\rightarrow \infty$) is given by
\be\label{eq:quadratic}
\eta = 1 - \frac{a}{2} - \sqrt{\left(\frac{a}{2}\right)^2 - b}.
\ee

For the cubic equations in tables~\ref{tab:pran34} and \ref{tab:pranzero} ($\delta^3 + a\delta^2 + b\delta + c = 0$), the shock solution is
\be\label{eq:cubic1}
\eta = 1 - \left(\frac{a}{3} + A + \frac{p}{A}\right),
\ee
where
$$
A \equiv \left(q + \sqrt{q^2 - p^3}\right)^\frac{1}{3},\;\;
p \equiv \left(\frac{a}{3}\right)^2 - \frac{b}{3}, \;\;
q \equiv \left(\frac{a}{3}\right)^3 - \frac{ab}{6} + \frac{c}{2}.
$$
For $R = 3/2$ and $n = 3/2$, the solution is given by expression (\ref{eq:cubic1}) for $f < f_c$, where $f_c$ is given in table~\ref{tab:bcp}, and by
\be\label{eq:cubic2}
\eta = 1 + 2\sqrt{p} \cos \left(\frac{\theta - 2\pi k}{3}\right) - \frac{a}{3},\;\; \theta \equiv \cos^{-1} \left(\frac{-q}{p^{3/2}}\right)
\ee
for $f > f_c$ (with $k = 0$). For $n = -2$, the solution is given by expression (\ref{eq:cubic2}) for $f < f_c$ (with $k = 1$), and there is a discontinuity at $f = f_c$ where the solution transitions from $2\eta_i - \eta_1$ to $\eta_1$ \citep{Zel'dovich02,Johnson13}. Evaluating expression (\ref{eq:cubic2}) can be problematic as $x \rightarrow \infty$ owing to the subtraction of two large numbers that are nearly equal. This can be seen in the panel $(b)$ of figures~\ref{fig:R43_R32} and \ref{fig:n43_n32}, where a glitch in the density appears near the post-shock region. The data for these plots (generated with NumPy) was noisy beyond this point and was replaced with post-shock values at infinity.

For the quartic equations in tables~\ref{tab:pran34} and \ref{tab:pranzero} ($\delta^4 + a\delta^3 + b\delta^2 + c\delta + d = 0$), the shock solution is
\be\label{eq:quartic}
\eta = 1 - \frac{a}{4} + \frac{k}{2}B - \frac{k}{2}\sqrt{-B^2 + 3r - s\frac{k}{B}}\,,
\ee
where
$$
A \equiv \left(q + \sqrt{q^2 - p^3}\right)^\frac{1}{3},\;\;B \equiv \sqrt{r + A + \frac{p}{A}}\,,
$$
$$
p \equiv \left(\frac{b}{3}\right)^2 - \frac{ac}{3} + \frac{4d}{3},\;\;
q \equiv \left(\frac{b}{3}\right)^3 - \frac{abc}{6} + \frac{a^2d}{2} + \frac{c^2}{2} - \frac{4bd}{3}\,,
$$
$$
r \equiv \left(\frac{a}{2}\right)^2 -\frac{2}{3}b,\;\;
s \equiv \frac{a^3}{4} - ab + 2c\,,
$$
and the value for $k$ is given in table~\ref{tab:bcp}. For $n = -3$, there is a discontinuity at $f = f_c$ where the solution transitions from $2\eta_i - \eta_1$ to $\eta_1$  \citep{Zel'dovich02,Johnson13}.

\setlength{\tabcolsep}{10pt}
\begin{table}
  \begin{center}
\def~{\hphantom{0}}
  \begin{tabular}{ccc}
      Solution & $k$ & $f_c$ \\[3pt]
      $R = 4/3$  & $1$ & $$ \\
      $R = 3/2$ & $0$ & $9/4$ \\
      $R = 4$  & $1$ & $$ \\
      $n = -3$  & $-1$ & $\left(\delta_1 \delta_i^3 - \delta_i^4\right)^{1/2}$ \\
       $n = -2$  & $1$ & $\left(\delta_i^3 - \delta_1 \delta_i^2\right)^{2/3}$ \\
       $n = 4/3$  & $1$ & $$ \\
       $n = 3/2$ & $0$ & $\left(27\delta_1/4\right)^2$ \\
       $n = 4$  & $1$ & $$ \\
       $n = \infty$ & $$ & $-\delta_1$ \\
  \end{tabular}
  \caption{Branches and critical points}
  \label{tab:bcp}
  \end{center}
\end{table}

The translational invariance of the equations allows one to multiply $f$ by any constant factor. To set the origin $x = 0$ at $\eta = \eta_o$, where $\eta_1 < \eta_o < 1$ but is otherwise arbitrary, multiply $f$ by a scale factor $S$, where $S$ is obtained from the relevant equation. For example, the equation for $R = 2$ with $f \rightarrow Sf$ is
$$
\delta^2 - Sf\left(\delta + 1/2\right) = 0.
$$
Since $f = 1$ at $x = 0$, this equation can be solved for $S$ to give
$$
S = \frac{\delta_o^2}{\delta_o + 1/2},
$$
where $\delta_o = \eta_o - 1$.

\bibliographystyle{jfm}

\bibliography{jfm_arxiv}

\end{document}